\documentclass[9pt]{extarticle}

\usepackage[a4paper,twocolumn,ignoreheadfoot,ignoremp,left=2cm,right=2cm,top=2cm,bottom=2.5cm,columnsep=1cm]{geometry}    

\usepackage[T1]{fontenc}
\usepackage[utf8]{inputenc}
\usepackage[swedish,english]{babel}

\usepackage{cmap}

\usepackage{abstract}
\usepackage{authblk}

\usepackage{graphicx}
\usepackage{float}

\PassOptionsToPackage{hyphens}{url}
\usepackage[hidelinks]{hyperref}

\usepackage{cite}

\usepackage{titleps}
\usepackage[font=small,labelfont=bf]{caption}

\usepackage{libertine}
\usepackage[largesc,looser,scaled=.92]{newpxtext}
\usepackage[scaled=.92]{newpxmath}
\usepackage{newtxtt}

\usepackage{microtype}

\usepackage[all]{nowidow}

\usepackage[absolute,overlay]{textpos}
\usepackage{pbox}

\hypersetup{
	hidelinks=true, 
	pdfinfo={
		Title={Addressing the Heterogeneity of Visualization in an Introductory PhD Course in the Swedish Context},
		Author={Kostiantyn Kucher, Niklas R{\"o}nnberg, Jonas L{\"o}wgren},
		Keywords={computing education; doctoral studies; visualization; human-computer interaction; computer graphics; media design},
  }
}

\begin{document}

\begin{textblock*}{80mm}(20mm,267.5mm)%
\tiny%
\noindent\textcopyright 2025 The Authors. 
This work is licensed under CC BY 4.0. To view a copy of this license, visit\\ \url{https://creativecommons.org/licenses/by/4.0/}
\end{textblock*}%

\title{
Addressing the Heterogeneity of Visualization in an \\ Introductory PhD Course in the Swedish Context}

\author{Kostiantyn Kucher\thanks{e-mail: kostiantyn.kucher@liu.se}}
\author{Niklas R{\"o}nnberg\thanks{e-mail: niklas.ronnberg@liu.se}}
\author{Jonas L{\"o}wgren\thanks{e-mail: jonas.lowgren@liu.se}}
\affil{Department of Science and Technology, Link{\"o}ping University, Norrk{\"o}ping, Sweden}

\date{August 12, 2025}

\twocolumn[
  \maketitle            
  \begin{onecolabstract}
    \noindent%
Visualization is a heterogeneous field, and this aspect is often reflected by the organizational structures at higher education institutions that academic researchers in visualization and related fields including computer graphics, human-computer interaction, and media design are typically affiliated with. 
It may thus be a challenge for new PhD students to grasp the fragmented structure of their new workplace, form collegial relations across the institution, and to build a coherent picture of the discipline as a whole. 
We report an attempt to address this challenge, in the form of an introductory course on the subject of Visualization Technology and Methodology for PhD students at the Division for Media and Information Technology, Linköping University, Sweden. 
We discuss the course design, including interactions with other doctoral education activities and field trips to multiple research groups and units within the division (ranging from scientific visualization and computer graphics to media design and visual communication). 
Lessons learned from the course preparation work as well as the first instance of the course offered during autumn term 2023 can be helpful to researchers and educators aiming to establish or improve similar doctoral courses. 

    \\
    \\
    \noindent \textbf{Keywords}: Computing education, doctoral studies, visualization, human-computer interaction, computer graphics, media design.
    \vspace{0.5cm}
  \end{onecolabstract}
]
\saythanks 

\enlargethispage{-0.8cm}

\section{Introduction}\label{sec:introduction}

The norms and practices of PhD education vary greatly across nations and regions, higher education institutions (HEIs) and their constituent parts, and, of course, disciplines and fields.
The onboarding process can thus be overwhelming for newly admitted PhD students, especially in heterogeneous environments that many researchers and educators in visualization and related fields find themselves affiliated with. 
However, preparing and informing the new PhD students of the official norms as well as local practices and traditions is a crucial step to ensure an engaged and productive activity at the respective organizational units. 
While the academic visualization community has previously engaged in the discussion of curricula, syllabi, and further pedagogical concerns relevant to first- and second-cycle higher education (e.g., the design of bachelor's or master's courses~\cite{Kerren2008Teaching,Kucher2021Project,Owen2013How}), the prior work on PhD education in visualization and related fields is surprisingly scarce. 
That is the \textbf{gap} this manuscript is aimed towards addressing.

In this paper, we report on the motivation, design, and outcomes of the first offering of an introductory course on the subject of Visualization Technology and Methodology (VTM)~\cite{GeneralStudyPlanVTM} for PhD students at the Division for Media and Information Technology, Linköping University, Sweden during autumn term 2023. 
This course, titled ``Introduction to PhD studies in VTM'', was aimed to complement the more general mandatory introductory course offered by the faculty (above the level of our division or institution)~\cite{Lith-Intro-Syllabus}. 
The unique focus of the proposed course lies in introducing the participants to disciplinary traditions in visualization and related fields, and furthermore, exposing the first- and second-year PhD students to the breadth and depth of research topics, methods, and activities across the units and groups within the division. 
Description of the course design, outcomes, and the discussion of lessons learned constitute the main \textbf{contributions} of this manuscript, while the intended target audience includes researchers and educators in visualization and related fields.

The rest of this manuscript is organized as follows: in the next section, we provide a brief overview of the prior work relevant to PhD education in visualization and related fields. 
In Section~\ref{sec:third-cycle-background}, we describe the background of the doctoral education in Sweden, as we acknowledge the existing differences with other educational systems within and especially outside of the European Union. 
We then describe the design of our course in Section~\ref{sec:course-design}, followed by the results of the first course instance offered during autumn 2023 in Section~\ref{sec:course-offering-results} and further reflections and discussions in Section~\ref{sec:discussion}. 
Finally, Section~\ref{sec:conclusions} concludes this paper.

\section{Prior Work on Doctoral Education in Visualization and Related Fields}\label{sec:related-work}

We can identify several areas of prior work relevant to the contents of this paper: 
first of all, we should mention the existing national and local regulatory information, including the Swedish Higher Education Ordinance~\cite{HigherEducationOrdinance} and the general study syllabus for VTM maintained for PhD education at our division~\cite{GeneralStudyPlanVTM}.  
While these documents set out overall aims and requirements for doctoral education, they do not specify concrete course-level goals or activities. 

Besides the formal regulations, we acknowledge the sources on pedagogy in higher education, especially PhD studies and supervision~\cite{Langs1994Doing,Wiener2003Supervising} (including the Swedish context in particular~\cite{Brodin2020Doctoral,SULF2015OnBeing}), as well as guides to the profession~\cite{AnderssonBurnett2022Beginner,SULF2020Expedition}, academic life~\cite{Fanghanel2011Being}, and research practices and ethics~\cite{VR2017Good} for PhD students. 
Some of these previous contributions have served as valuable sources for our introductory PhD course, however, they do not provide any discipline-specific guidance. 

Focusing on the prior work within visualization and related fields, we should mention the existing publications focusing on visualization literacy~\cite{Firat2022Interactive} concerns both outside and inside of formal education contexts, such as studies conducted with science museum visitors~\cite{Boerner2016Investigating}, Amazon Mechanical Turk crowd-sourced participants~\cite{Tanahashi2016AStudy}, biochemistry students~\cite{Schoenborn2006Importance}, or business students~\cite{Grammel2010How}. 
While the respective studies investigate the role, level, and assessment techniques for visualization literacy, they do not specifically target doctoral students in visualization and related fields. 
Similarly, recent studies on visualization onboarding~\cite{Stoiber2022Comparative,Stoiber2023Visualization} and visualization interaction strategies in informal learning environments such as museums and science centers~\cite{Yu2025Revealing} provide suggestions for accommodating visualization novices in particular workflows and tools. 
As Sch{\"o}nborn and Besan{\c{c}}on argue~\cite{Schoenborn2024What}, the perspectives and methodologies offered by educational science itself may be applicable within visualization research. 
While all of these studies are valuable, they do not address the problem of introducing novice doctoral students to the discipline of visualization. 

Considering the formal education settings, we can also highlight the prior work describing the course design for first- and second-cycle higher education in information visualization and visual analytics~\cite{Kerren2008Teaching,Kerren2013Information,Kucher2021Project,Owen2013How}, including visualization for multi-/interdisciplinary and non-technical programmes~\cite{Jaenicke2019AVisualization,Jaenicke2020Teaching}. 
Furthermore, discussions of well-established and novel teaching methodologies~\cite{Bach2023Visualization,Syeda2020Design}, concerns~\cite{Rodrigues2021Computer}, reflections~\cite{Aerts2022Me-ifestos}, and community efforts~\cite{Diehl2021VisGuided} can be found in the literature in relation to visualization and related fields. 

However, very little prior work focusing on PhD-level education concerns within our fields is available: this includes the peer-reviewed publications~\cite{Laramee2011How,McNabb2019How}, book chapters~\cite{Munzner2008Process}, and blog posts~\cite{Elmqvist2015How,Elmqvist2019Writing} from the members of our community on the concerns related to research paper analysis and writing as well as the peer review process; 
arguably, the prior work on the ethical concerns relevant to visualization research~\cite{Correll2019Ethical} also fits in this list. 
With the current manuscript, we aim to address this gap in the literature on doctoral education in visualization.

\section{Background: Doctoral Education in Sweden}\label{sec:third-cycle-background}
In this study, we focus on the Swedish context, where PhD students typically take on several roles at HEIs, including those of professional researchers, teachers, and actual students. 
The national regulations for PhD education are set by the Higher Education Ordinance~\cite{HigherEducationOrdinance}, with the learning objectives for the complete degree divided into three categories: Knowledge and understanding, Skills and abilities, and Judgement and approach. 
Knowledge and understanding consider demonstration of a broad knowledge within the research field at large and an in-depth and current specialist knowledge with scientific methods within this research field. Skills and abilities concern, among other aspects, demonstration of a capacity for scientific analysis, the ability to make significant contributions to the research field, and the skill to present research findings in both orally and in writing. Judgement and approach consider the demonstration of intellectual independence and the insight into the potential and limitations of the research performed.
PhD students demonstrate the fulfilment of the objectives via the doctoral thesis/dissertation (often based on a number of peer-reviewed publications authored by the respective PhD candidate) as well as coursework. 
The latter typically ranges between 30--120 European Credit Transfer and Accumulation System (ECTS) credits and includes both mandatory and selective courses. 
The complete PhD degree in Sweden corresponds to 240 ECTS credits, i.e., four years of full-time activity (usually complemented by some teaching activities or other departmental duties, as the majority of PhD students are employed at the respective institutions). 
Discipline-specific and local regulations for PhD education, including the coursework scope and mandatory courses, are specified in \emph{general} study syllabi designed and issued at HEIs.
Each PhD student (alongside their main supervisor and co-supervisors) is also required to maintain and fulfill an \emph{individual} study plan.  
There are typically few mandatory courses included in the general study syllabi, such as philosophy of science and research ethics---and such mandatory courses are typically offered in a rather centralized way at the respective HEIs. 

For example, the mandatory introductory PhD course offered at the level of the Faculty of Science and Engineering at our university (above the level of our division) comprises the following four modules~\cite{Lith-Intro-Syllabus}: introduction to scientific methodology and basic skills for PhD studies; sustainable research and society; gender and equality; and research ethics. 
One downside of relying on rather general introductory courses designed for wider PhD student audiences is that discipline- or field-specific concerns and conventions are typically missing from the respective courses.

\section{Doctoral Course Design}\label{sec:course-design}

\begin{figure*}[ht!]
\centering
    \includegraphics[width=0.995\linewidth]{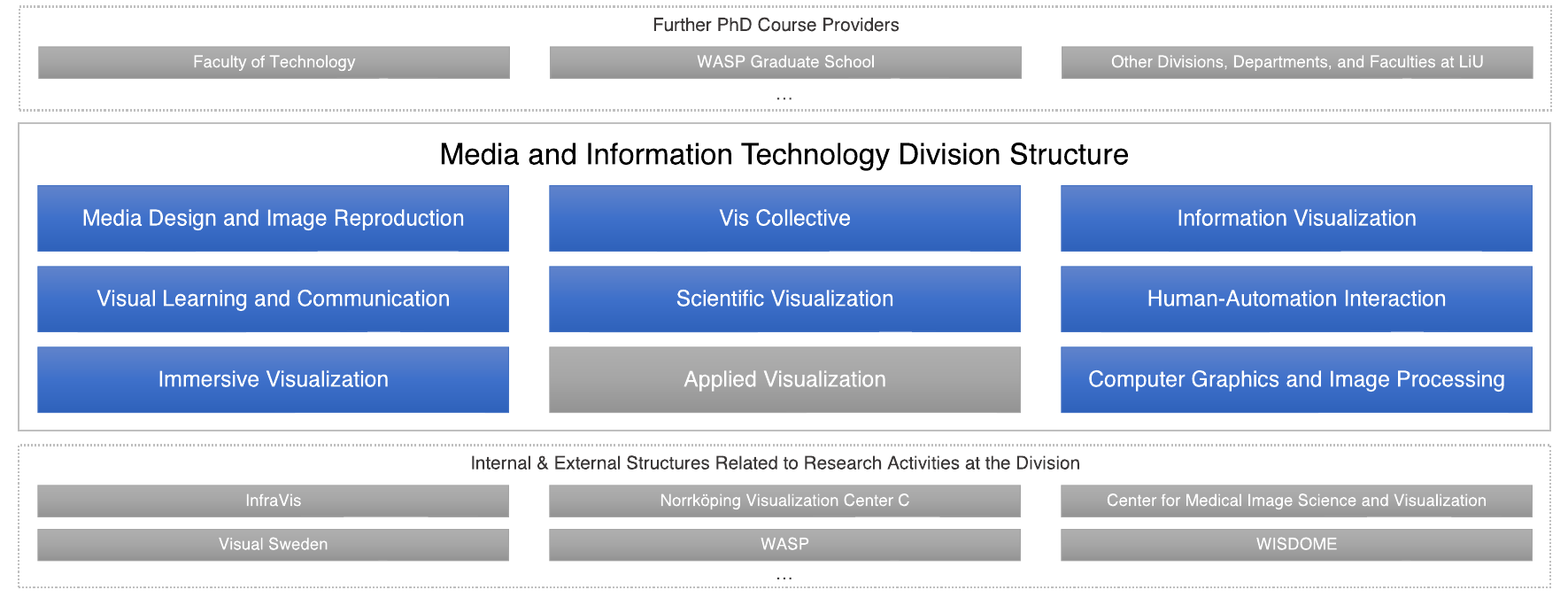}
     \caption{Structure of the research units and groups at the Division for Media and Information Technology (MIT), Linköping University, Sweden at the time of first course offering during autumn term 2023. 
     The units and groups pursuing PhD education are highlighted in blue.  
     The structures within the division also collaborate with a number of other organizational structures within the university as well as beyond (such as the external WASP research funding program or the WISDOME initiative on research and education in 3D dome environments), often resulting in activities and contributions relevant to PhD education. 
     Furthermore, PhD education at the division as well as the course design described in this manuscript are affected by other existing regulations and commitments, such as mandatory PhD courses offered by the faculty~\cite{Lith-Intro-Syllabus} or external PhD schools (e.g., the WASP Graduate School funded by the respective program) that some of the PhD students at MIT are affiliated with.}%
     \label{fig:mit-structure}%
\end{figure*}

Our motivation for designing this course for first- and second-year PhD students at the Division for Media and Information Technology should be explained in the context of the diversity of fields and topics addressed at our division, reflected in the organizational structure presented in Figure~\ref{fig:mit-structure}. 
The research units and groups within the division pursue research (and applied work) across several related fields and subfields: visualization and human-computer interaction (from scientific and immersive visualization to information visualization, visual analytics, sonification, visualization in a broader context, and human-AI/automation teaming); media design; image reproduction and processing; computer graphics; and visual learning and communication. 
One of the practical aims of the proposed course was thus to bring PhD students from different units and groups together.  
However, the deeper aim was to build the foundations for an understanding of the (heterogeneous) discipline as a whole, including the common underlying concerns and methods (potentially including even further topics and fields, such as mathematics, software engineering, and perception and cognition, for instance).

Our course was thus designed to complement the introductory PhD course offered by the faculty (see Section~\ref{sec:third-cycle-background}). 
As the intended learning outcomes, we have stated that the PhD students are expected to attain the following on successful course completion:
\begin{itemize}
\item A general understanding of the formal, practical, and social structures underpinning PhD studies in VTM, and a specific understanding of the personal implications of those structures.
\item A general understanding of the norms of scientific practice in VTM research, and a specific understanding of what those norms imply for the PhD student’s own PhD studies.
\item A general understanding of the scope of the research subject of VTM as practiced in our division, and a specific understanding of how own research interests and plans relate to other research in the division.
\end{itemize}

For practical reasons, the course was organized in three independent parts, each focusing on one of the learning outcomes listed above and resulting in 2 ECTS credits (see Figure~\ref{fig:course-timeline}). 

The first part of the course focuses on roles, responsibilities, and paths for PhD students in VTM. 
The topics addressed here include practical and administrative fundamentals; the multiple PhD student roles (a course-taking student, a researcher in training, and typically, a colleague, an employee, and a teaching assistant); national learning goals for the PhD~\cite{HigherEducationOrdinance}; the progression of becoming a researcher~\cite{SULF2020Expedition,AnderssonBurnett2022Beginner}, the supervisor-PhD student relation~\cite{SULF2015OnBeing}, work life and time management.
The second part focuses on norms of scientific practice in VTM, including criteria for academic research and researchers, publication practices and ethics, among others~\cite{VR2017Good,ACM-CoE,IEEE-CoE}. 
The teaching and learning activities for these two course parts mainly consist of seminars that typically rely on essay preparations, requiring the course participants to review the existing sources and discuss the respective concerns with their PhD supervisors. 
Finally, the third part of the course focuses on the scope of VTM, including the ongoing and planned research in the units and research groups of the division. 
The activities here include ``field trips'' to the respective units and groups, typically involving presentations from the respective researchers, demonstrations of their work, and hands-on activities with the respective hardware and software, where applicable. 
The course is mainly intended for on-site participation, but remote attendance (e.g., via Zoom) is permitted.

Each of the course parts is examined through an individual assignment requiring synthesis and reflection on the individual implications of more general knowledge. 
Part 1 is examined through a written manifesto or personal plan, setting goals and identifying concerns for the PhD student’s years ahead. 
Part 2 is examined through a paper on a topic of particular interest within norms of scientific practice, how the topic is approached in the different research units of the division, and what the personal implications are for the years ahead.
Part 3 is examined through a paper identifying a small number of topical connections between the PhD student’s own plans and other ongoing or planned research within the division, including possible collaborations.
The final papers are also to be discussed (as either brief presentations or panel discussions) at the seminars concluding each course part.

As the course materials, we mainly rely on the sources mentioned above, especially for the first and second parts of the course, while the third one relies on the publications suggested by the respective research groups and units. 
The course was designed to be offered in English due to the large number of doctoral students and researchers with an international background at our division, but also the reliance on the research contributions and prior work in visualization and related fields being typically published at international venues.

\section{First Course Offering Results}\label{sec:course-offering-results}

\begin{figure*}[ht!]
 \centering
 \includegraphics[width=0.995\linewidth]{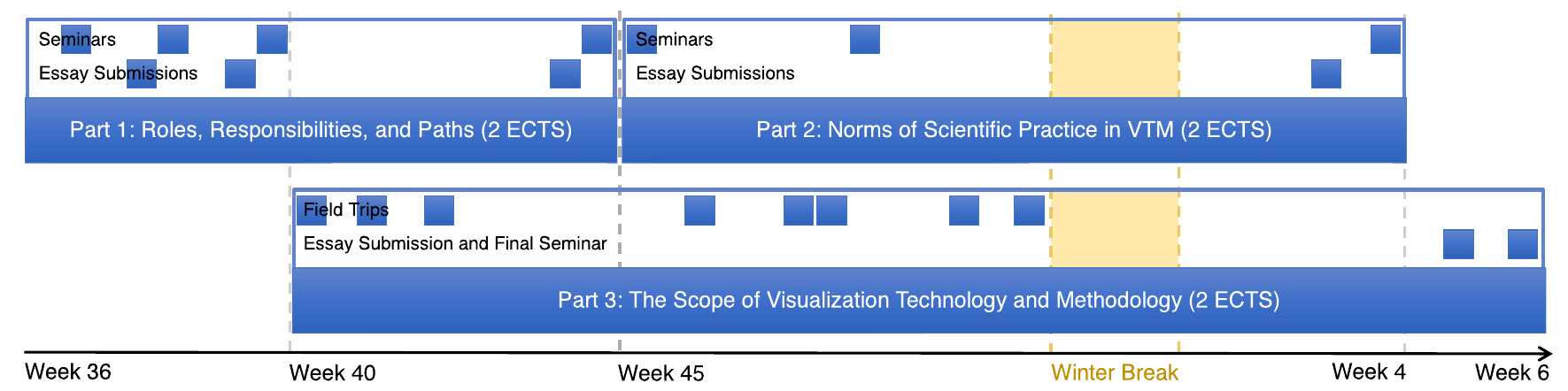}
  \caption{The timeline of our introductory PhD course on the subject of Visualization Technology and Methodology (VTM), as offered for the first time during the autumn term 2023. 
  Each part of the course corresponds to 2 European Credit Transfer and Accumulation System (ECTS) credits. 
  The participants are expected to take this course alongside other mandatory and/or selective PhD courses as well as research and departmental duties, as customary in the Swedish higher education system.}
   \label{fig:course-timeline}
\end{figure*}

The first course offering took place during the autumn term 2023 (see Figure~\ref{fig:course-timeline} for the timeline), with the course introduction and Part 1 kick-off at the beginning of September 2023. 
As the three course parts were offered independently, the number of course participants for each part was different, with more PhD students in their second year registering for Part 2 and Part 3, for instance. 
The participants were quite diverse in terms of the affiliated research group/unit, gender, origin, educational background (technical vs non-technical disciplines, for instance), and even academic age (with some participants just being admitted around the course start date). 
From the teachers' perspective, the organization of the course offering was synchronized and carried out with one coordinator / active teacher per course part (among the authors of this manuscript), while many colleagues from the division were involved as invited presenters (especially for Part 3). 
In the following subsections, we summarize the outcomes of each of the three parts.

\subsection{Part 1: Roles, Responsibilities, and Paths}
    
The first part of the course consisted of four sessions of two hours each. 
These sessions were held during September and at the beginning of November 2023.
Seven PhD students took part in this part of the course. 
Each session (apart from the last one) had an assignment that was designed to encourage the PhD students to discuss different topics with their supervisor(s). 
The following session began with a presentation and discussion of the assignment.

This first session of the first part of the course focused on what it is to be a PhD student within the division of Media and Information Technology, with information and discussions on 1) the general study plan for PhD studies in the field of Visualization and Media Technology, 2) the individual study plan in relation to the general study plan and the learning criteria and assessments of PhD studies in Sweden, 3) course requirements related to the subject as in number of course credits needed and faculty course requirements, 4) the format of the thesis (monograph or compilation thesis), 5) supervision in terms of both practical and interpersonal aspects. 
The task related to this session was for the PhD students to discuss third-cycle level courses (i.e., PhD level courses) with their supervisor(s), what PhD courses to choose, why choosing these courses, and when during the PhD time these courses would be useful. 
The PhD students were asked to consider PhD-level (third-cycle) courses, as well as master-level courses, seminars and reading groups, and possible individual reading courses.

The second session considered the roles of a PhD student at the division in terms of being a student and learning how to become an independent researcher, being an employee and part of a larger organization in the division, in the department, and at the university, and finally, being a member of a research project.
This session also included a discussion on the PhD learning outcomes and degree objectives as specified in the general and the individual study plans. 
Finally, members from HR also took part of this session, giving further information about rules, regulations, and rights of the PhD students.
The task for this session was to discuss with their supervisor(s) how the different learning objectives as specified in the individual study plan can be achieved and fulfilled.

The third session focused on responsibilities and the how these responsibilities might change between the supervisor(s) and the PhD student during the years, but the session also discussed the amount of work hours that is reasonable to spend working, who the PhD students should turn to for conversations if necessary, and about the relationship to the supervisor. 
This discussion led to a conversation about the supervision process, the differences between supervisor and co-supervisor, and about academic mentorship, the academic culture, co-authorship, and collegial support. 
After these discussions, an appendix to the individual study plans was introduced. 
This appendix presents a few different scenarios or topics that the PhD student and the supervisor(s) discuss and mark who is most responsible. 
The reason for using this appendix is to promote discussion and consensus between PhD student and supervisor(s). 
The task for this session was to write a text, a manifesto, that set the goals of and identifying concerns for the years ahead of the PhD students. 
The text was supposed to take the standpoint in the learning outcomes stated in the individual study plan a PhD education in terms of knowledge and understanding, competence and skills, and judgement and approach.
The PhD students got some tentative questions to use for their reflections such as: \textit{What are the goals with your PhD, and how will you achieve these? What is your responsibility as a PhD student, and what are the responsibility of your supervisor(s)? What are the concerns (in terms of research, teaching, courses, publications, work and life balance), and how can you address these concerns? What courses are relevant for you to take, what courses are available and what courses would be great to find? To what extent would teaching be of interest and use for you, in what courses, and who has the authority to decide if you should be teaching or not? What conferences would be of use for you to visit, and do you need to present you own research at these conferences? What are the relevant publication venues for your future research, and how do find these?}

The fourth session consisted of the PhD presenting and discussing these texts. 
By sharing their plans, thoughts, and worries for the coming years, the PhD students realized that they shared many of these concerns and they were involved in interesting conversations and discussions.

In the manifestos written by the PhD students, there were some common thoughts and concerns for the future. These concerns mainly considered:
\begin{enumerate}
    \item Work-life balance, and the importance of establishing boundaries and assessment of the workload;
    \item Managing teaching time in relation to research time, as teaching is meritorious for a future career in academia while also requiring a fair amount of time and engagement;
    \item Finding and finishing relevant courses at third-cycle level and getting/having the right competence that is needed for the research, or finding the right collaborator(s) to join a research project;
    \item Attending conferences, submitting and being accepted to relevant conferences and finding funding for attendance;
    \item Figuring out a sound publication strategy that allows for timely publication in relevant venues;
    \item Concerns for after what to do after successfully defending the thesis, staying in academia or leaving for commercial business, and how a doctoral degree stands up to the competition; and
    \item The novelty, and the width and depth, of the research topic, and the research design and the emergence of relevant sub-studies.
\end{enumerate}

\noindent One PhD student stated in the end of the document that the \textit{``manifesto serves as a guiding document, evolving with the dynamic nature of academia, as I contribute meaningfully to the scholarly discourse and prepare for a future of continued research and innovation''}. 
While another PhD student wrote that by \textit{``communicating openly with our supervisors, we aim to work together to achieve our goals and make meaningful contributions to our field. Although challenges lie ahead, our dedication, determination, and ongoing commitment to learning will empower us to succeed in this academic journey''}.

\subsection{Part 2: Norms of Scientific Practice in VTM}

The second part followed the first sequentially, focusing on the norms of scientific practice in visualization research and specifically in the heterogeneous research context of our division. 
Nine PhD students completed the second part, which started in early November 2023 and concluded in mid-January 2024.

The first session was devoted to the ethical rules for academic conduct provided by the Swedish Research Council~\cite{VR2017Good}:
{\itshape
\begin{itemize}
    \item You shall tell the truth about your research.
    \item You shall consciously review and report the basic premises of your studies.
    \item You shall openly account for your methods and results.
    \item You shall openly account for your commercial interests and other associations.
    \item You shall not make unauthorized use of the research results of others.
    \item You shall keep your research organised, for example through documentation and filing.
    \item You shall strive to conduct your research without doing harm to people, animals or the environment.
    \item You shall be fair in your judgment of others' research.
\end{itemize}
}
\noindent The Swedish Research Council is considered the most authoritative funding agency in Sweden for STEM subjects, and their ethical guidelines~\cite{VR2017Good} are widely used and referred to. 
During the session, we addressed each rule in turn and the PhD students provided examples, questions, and dilemmas that were discussed in order to contextualize the general rules into our everyday research environment. 
The discussions were live-documented in a collaborative whiteboard. 
The first rule, for example, stimulated an important discussion on the tradeoff between academic integrity and bibliometric performance. Another example 
was plagiarism, particularly in the age of generative AI.

The homework after the first session consisted of a deep dive into the specific subject of joint authorship, as it was deemed to be one of the most salient aspects of research ethics for young PhD students. 
The supervisors in the division had prepared a memo on joint authorship practices in our field, summarizing core guidelines such as the Vancouver Recommendations~\cite{Vancouver} and discussing common concerns such as ordering author names and determining whether a project contributor should also be an author. Students read the memo and discussed joint authorship with their respective supervisors, focusing on the practices in their specific units.

\begin{figure*}[ht!]
 \centering
 \includegraphics[width=0.995\linewidth]{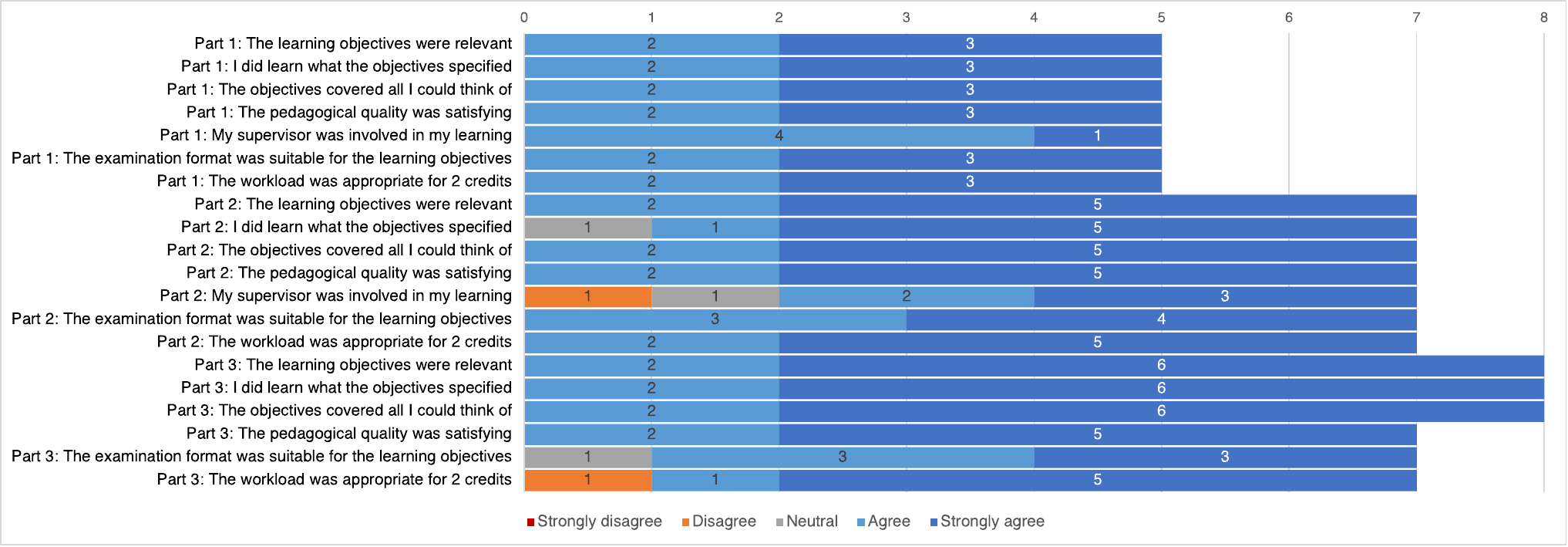}
  \caption{Course evaluation results for the autumn term 2023. 
  Replies concerning Parts 1--3 of the course were provided by 5, 7, and 8 respondents, respectively.}
   \label{fig:course-evaluation-results-2023}
\end{figure*}

The second session started with reporting and discussing the homework assignment. 
Several of the students reflected on common engineering-research dilemmas such as publishing a piece of research that has required significant technical work, and what the research engineers doing the technical work in such projects would need to do to be included as authors. 
Another potential dilemma concerned dealing with politically motivated gift authorships.
The next part of the second session provided an introduction to the recommended readings from the syllabus, and the session closed with introducing the examination assignment.

As examination, the students wrote individual papers on topics of particular interest within the norms of scientific practice. 
Student papers were organized into themes and the part concluded with an open division seminar where the students formed a panel and the discussion was moderated by the teacher responsible for the course part. 
The two overall themes for the panel discussion at the seminar were ``The data and methods we use'' and ``Research for the common good''.  Panelists as well as audience members were engaged in the discussion, unpacking topics such as academic transparency, trading off truthfulness and academic salesmanship, developing working systems for others to use, and engaging with societal challenges as part of PhD research.
    
\subsection{Part 3: Scope of VTM}

The third part of the course was designed over the complete term. 
Overall, twelve participants from five research groups/units engaged in this course part. 

Field trips to all eight research groups and units involved in PhD education at our division (see the blocks highlighted in blue in Figure~\ref{fig:mit-structure}) took place. 
The duration and contents of the respective field trips varied greatly, ranging from demos of specialized hardware related to 3D printing (Media Design and Image Reproduction), a 3D dome (Immersive Visualization + Applied Visualization), and an air-traffic control simulator lab (Human-Automation Interaction) to more conventional presentations given by the leaders and participants of the respective groups/units. 
The course participants themselves belonged to some of these research units and groups, and were thus involved in the ``hosting'' activities in some cases. 
This further highlights the duality of the PhD students' roles as researchers and actual students, but also opens up opportunities for collaborations initiated by the PhD students through such courses. 

This course part concluded with the final essay being submitted in early February 2024, 
followed by a panel discussion 
with a format similar to Part 2 described above. 
The course participants and the members of the division attending this seminar discussed the common themes and specific concerns raised in the essays. 
These included the acquaintance with the research topics and methods at the participants' own and other units/groups; finding common ground and potential forms of collaboration for the future; and potentially taking the lessons learned within this course to the rest of the division staff beyond the first- and second-year PhD students.

\subsection{Course Evaluation Results}

In order to collect further feedback from course participants and improve future course offerings, we reached out to the participants after the course finish and asked them to fill out a questionnaire with both closed (5-point Likert scale) and open-ended questions. 
The participation in course evaluation was voluntary and anonymous. 
We have received responses from 8 participants, and the results for close-ended questions are summarized in Figure~\ref{fig:course-evaluation-results-2023}. 
The overall feedback from the participants was overwhelmingly positive, with comments such as ``\textit{enjoyed the open discussions where everyone shared their thoughts and understanding}'' for Part 1, ``\textit{very good learning experience}'' for Part 2, and ``\textit{It gave me an insight of different research that is happening in MIT division, so that I know who should I reach out to in case if I need any help in my research. It gave me a platform to present my research project, so that others are aware of what I'm working on and can reach out to me if they want to collaborate. Finally, it gave me an opportunity to learn new things that are within and beyond the scope of my research.}'' for Part 3. 
Regarding the course in general, the participants left comments such as ``\textit{a well-designed course that benefits both students and the division; a pleasant experience to explore different research directions}''; ``\textit{I had an amazing experience taking this course and would recommend this course to new PhD students so that they will be aware of what is going on in the MIT division and won’t feel excluded.}''; and ``\textit{The course was excellent for gaining insights into various aspects essential for PhD students. I highly recommend it to anyone embarking on their PhD journey.}''. 

Besides expressing the overall appreciation, the participants also provided some constructive suggestions for course improvement. 
For example, regarding the panel discussion format for Parts 2 and 3, one of the participants wrote: ``\textit{The panelist approach for examination was interesting and seemed like it could work well. However, I don't know how useful it actually was for us PhD students. It felt very weird to be on a `panel' before a group of more experienced colleagues when those were the people whose thoughts and opinions are considered more interesting.}'' 
The participants also made some remarks about the scheduling of field trips in Part 3, as it was difficult for some participants to attend all eight separate sessions (and they suggested switching to an intense series of workshops during a dedicated week instead, for instance).
Regarding the discussions held during the individual field trips vs the final panel discussion in Part 3, one participant stated ``\textit{I wish there could be a short moment to reflect and brainstorm within each session rather than all together in the final seminar.}''
This suggestion was also mirrored by a comment from another participant: ``\textit{maybe a small brainstorming session can be added at the end of each field trip}''. 
While the format of each field trip was defined by the respective group/unit, the feedback from the course participants has provided us with directions for improvements for future course offerings.

\section{Discussion and Reflection}\label{sec:discussion}

\paragraph*{Course design and dependencies:} 
The diversity of the topics within our division as well as additional constraints related to the mandatory courses taken by all/some of the intended course participants (see Figure~\ref{fig:mit-structure}) were definitely a challenge for us at the course design stage. 
At the same time, this heterogeneity was embedded in the topics discussed with course participants (and by course participants with their supervisors at the respective research groups and units) and presented during the field trips by other division staff members. 
For the last part of the course, the diversity of the topics also meant that we would not be able to rely only on the expertise of a small number of teachers closely associated with the course. 
Wile the logistics of such field trips required some planning and compromise, with the help and commitment of the colleagues from our division, the respective course part was carried out successfully.

\paragraph*{Course participants' involvement:} 
During the final presentations of the first part of the course, the PhD students shared their plans and concerns written in the manifesto. 
To the examiner, they were surprisingly open in their reflections for the coming years, sharing thoughts and doubts with each other. 
The students also supported each other in these thoughts and provided feedback and possible ideas to mitigate forthcoming obstacles and challenges. 
This openness was probably made possible thanks to the students themselves, but also as this part of the course focused on giving the students the basis for reflection on these topics and what it means to be a PhD student within the division.
In the presentations of the manifestos, there were some overlaps between the reflections and thoughts presented. 
This was experienced as something positive. 
The PhD students realized that they were not alone in their concerns for the coming years, and that they could support each other and discuss these issues.
In that way, they got the realization that they are not alone in being a PhD student but part of a larger group of supervisors, senior colleagues, administrative staff, and fellow PhD students.

\paragraph*{Cooperation and collaboration from the PhD students' perspective:}
As discussed at the final seminar for Part 3, crossdisciplinary collaboration is a broad spectrum ranging from inspirational conversations and serendipitous discoveries all the way up to joint projects and publications. 
The key for an institution that wants to stimulate crossdisciplinary collaboration is then to provide arenas facilitating crossdisciplinary exchanges, accommodating the whole range mentioned above. 
This could include recurrent seminars, informal gatherings, structured information and demo sessions, shared technical resources, mutually accessible labs, etc.
Some concerns raised by both course participants and senior staff members attending the seminar were also related to the risks of forcing collaborations or focusing only on the interests of one party---and it seemed that a common understanding of avoiding such risks was shared by the attendees. 

\paragraph*{Follow-up improvements for the autumn 2024 course offering:}
The lessons learned from the initial 2023 course offering led us to slight updates in the course design, formalized in the updated syllabi as three separate PhD courses in 2024~\cite{IntroVTM-Part1-Syllabus,IntroVTM-Part2-Syllabus,IntroVTM-Part3-Syllabus}. 
While we still encourage all of the newly admitted PhD students (and their supervisors) at our division to take all three course parts, the updated design offers flexibility in case of their limited time or availability. 
Furthermore, since organization of field trips to a number of research groups and units for Part 3 requires a lot of resources, we have introduced a scaled-down version of this course part that better accommodates smaller audiences (e.g., we have a smaller group of 3 course participants during the autumn 2024 course offering). 
In this version, we offer the previously collected materials from various groups and units to the participants for in-depth self-studies. 
Multiple field trips are replaced by a single full-day workshop with presentations from the respective groups/units, including the perspectives of PhD candidates and postdoctoral researchers. 
However, in order to provide the participants who complete this scaled-down version of Part 3 with complete experience, we intend to offer full-scale field trips every several years and invite the former participants to such events.
Our advice for educators designing similar courses in this regard is to consider the potential variations in the number of PhD students admitted to the course and the implications for the respective course activities and resources required. 

\paragraph*{Limitations:}
The design of the course described in this study was strongly affected by the current local, national (Swedish), and international (European) regulations and academic traditions. 
This naturally provides a risk for generalizability of our findings that we acknowledge; 
however, we believe that the lessons learned from our experiences could be applicable even for less formal training activities within doctoral education, e.g., within educational systems that do not include formal courses as part of PhD studies. 
The course design and outcomes described here are also based on limited data with respect to the temporal dimension, and thus reflections on the outcomes of future course offerings should be considered part of future work.

\section{Conclusions}\label{sec:conclusions}
In this manuscript, we have described the introductory PhD-level course titled ``Introduction to PhD studies in VTM'', designed for the subject area of Visualization Technology and Methodology at Linköping University, Sweden. 
We have motivated and discussed the course design, the results of the first course offering during the autumn term 2023, and lessons learned from these experiences. 

As part of the future work regarding the course maintenance and updates, we intend to consider further relevant course materials, such as discussions of peer-review activities in the visualization and HCI communities~\cite{Wilson2022How}. 
Furthermore, we intend to connect this course to the future methodology and academic writing courses to be developed within our division with the context of visualization and related fields in mind. 
Finally, we intend to (and urge the academic community) to share the pedagogical experiences and novel approaches related to PhD-level education in visualization and related fields to address the currently existing gaps in the literature.

\section*{Acknowledgments}
We would like to express our gratitude to the course participants and staff members at the Division for Media and Information Technology, Linköping University---this course would not exist and could not be offered without their contributions.


\bibliographystyle{abbrv-doi}

{
\small
\bibliography{bibliography}
}
\end{document}